\documentclass[preprint,aps,showpacs]{revtex4}
\usepackage{amssymb}

\usepackage{amsmath}
\usepackage{graphicx}
\usepackage{bm}
\usepackage{latexsym}

\begin{document}

\title{Density of states of a two dimensional XY model from Wang-Landau algorithm}
\author{ Jun Xu, H. R. Ma\\
 Institute of Theoretical Physics, Shanghai Jiao Tong University,\\
 Shanghai 200240, China}
\date{}

\begin{abstract}

Using Wang-landau algorithm combined with analytic method, the density of states
of  two dimensional XY model on  square lattices of sizes $16\times16$,
$24\times24$ and $32\times32$ is accurately calculated. Thermodynamic quantities,
such as internal energy, free energy, entropy and specific heat are obtained from
the resulted density of states by numerical integration. From the entropy curve
symptoms of phase transition is observed. A general method of calculation of the
density of states of continuous models by simulation combined with analytical
method is proposed.
\end{abstract}
\maketitle

\section{INTRODUCTION}

The XY model has been widely used in description of many physical
systems and in the studies of phase transitions and other related
problems. The two dimensional XY model is especially interested
in the studies of the KT transition.  The model has been
extensively studied in the past decades with many different methods,
while the density of
states(DOS) is yet not determined because of the complexity of the
model. R. Toral and his group computed the DOS of one dimensional XY model
and got some scaling results\cite{{RRA2002PA},{R2004JSP}}.
 However, the two
dimensional XY model is more interesting and valuable. It is famous for its
quasi-long range order, vortex pairs and KT phase transition
\cite{JV1979PRB}. It will be helpful for people to understand the
nature of the model and study in more details of the KT phase
transition if the DOS of the two dimensional XY model is calculated.

The model can be described by the Hamiltonian:
\begin{equation}\label{Ham1}
H=-J{\sum_{<ij>}}{\vec{S}_i}{\cdot}{\vec{S}_j}
\end{equation}
where $J$ is the coupling constant, and ${\vec{S}}$ is a unit vector(the spin) in
a plane. Here $i$ or $j$ is the lattice site on a plane, and the summation is
over all nearest neighbor pairs. Different from the Ising model, the XY model is
a continuous model in which the energy changes continuously. Thus the total
number of states of the whole system is infinite, this property gives rise to the
difficulties in determining the DOS near the highest and lowest energy level, as
we will explain in detail later.

Recently many kinds of simulation algorithms used in the  calculation of the DOS
have been proposed, in this study we  use Wang-Landau algorithm to calculate the
DOS of a two dimensional XY model\cite{{WL2001PRL},{WL2001PRE}}. The fast
convergence,  stability and easy to implement of the method makes it the first
choice in a DOS calculation, however, it has some shortcomings in the accuracy of
calculation as pointed out by many researchers\cite{CR2001PRE}. In the case of
the two dimensional XY model and other continuous models alike, the algorithm
usually fails to give good results of DOS close the the boundaries of the
spectrum. Fortunately, on the spectrum boundaries, there are usually analytical
results of the DOS available or may be obtained in less effort. By combining the
analytical results close to spectrum boundaries with simulation, better results
of the DOS can be obtained. In the following we show this process using the two
dimensional XY model.

\section{CALCULATIONAL PROCEDURE}
\label{sect2}

\subsection{the density of states}
In order to use the Wang-Landau algorithm to a continuous model, we divide the
whole energy range to some sub-intervals and fine divide each sub-interval into
many small intervals. Every small interval is regarded as an energy level and
represented by its middle energy. Then the Wang-Landau algorithm can be used to
calculate the relative DOS of each sub-intervals and then the DOS of the whole
range can be obtained by smooth join the DOS on each sub-intervals. It should be
noted that the DOS calculated this way is the relative DOS of the system, which
contains an arbitrary scaling constant. In the case of a discrete model like the
Ising model, both of the total number of states and the number of the states of
the lowest energy level are known and the scaling factor can easily be determined
by using either of the known conditions. In the case of a continuous model, one
has no knowledge of the DOS at any point so that it is harder to get the scaling
factor directly. In fact, the situation is even worse. In the XY model, the DOS
in the middle range of the spectrum do not change much so that the discretization
will not introduce much error to the calculation of DOS. On the other hand, the
changes of the DOS close to the spectrum boundary are very large, so accuracy
calculation of the DOS close to the spectrum boundary is impossible by simply
divide the small intervals finer and finer. This difficulty  is bypassed by the
observation that in many cases of the continuous models the analytical expression
of DOS close to the spectrum boundaries are known. In this case we can use the
known expression for the DOS close to the the spectrum boundaries and join them
smoothly to the simulated relative DOS on the middle to get the DOS of the
system.

The Hamiltonian of a two dimensional square lattice XY model can be written as:

\begin{equation}\label{Ham2}
H=-{\sum_{<ij>}}{\cos}({\theta_i}-{\theta_j}).
\end{equation}
Here we set $(J=1)$ for simplicity, and $\theta_i$ is the angel between
$\vec{S}_i$ and a reference direction in the plane. The ground state of the model
is the ferromagnetic ordered state where all $\vec S_{i}$ are parallel.  Close to
the ground state, each spin may deviate from the alignment direction by thermal
agitation. The deviation is small at low temperature, so that it can be
approximated as:
\begin{equation}\label{Ham3}
H=-2N+{\frac{1}{2}}{\sum_{<ij>}}({\theta_i}-{\theta_j})^2
=-2N+{\frac{1}{2}}{\sum_{\vec{k}}}{\vert}{\theta(\vec{k})}{\vert}^2f(\vec{k})
\end{equation}
Where $N$ is the total number of the square lattice sites
$(N=L{\times}L)$, and $\vec{k}$ is the wave vector in the plane,
$\theta(\vec{k})$ is the Fourier component of $\theta_i$ and
\[
f(\vec{k})={\sum_{\vec{a}}}(1-{\cos}(\vec{k}{\cdot}\vec{a})).
\]
Here $\vec{a}$ is the base vector of the squre lattice. The partition function
of the system is
\begin{equation}\label{pf1}
Z(\beta)={\int^{-2N}_{2N}}g(E)e^{{-\beta}E}dE
={\int^{4N}_0}g(E-2N)e^{-\beta(E-2N)}dE{\approx}e^{2{\beta}N}
{\int^{\infty}_0}g{\prime}(E)e^{-{\beta}E}dE
\end{equation}
Where $g{\prime}(E)=g(E-2N)$. From equation (\ref{pf1}) we see that
 $Z(\beta)e^{-2{\beta}N}$ is the
Laplace transformation of $g{\prime}(E)$. Here $\beta=1/T$ is the
inverse of temperature (we use a unit system where Boltzmann constant $k_B =1$).
From equation (\ref{pf1}) it is clear that the DOS $g'(E)$ can be obtained
if the partition function $Z(\beta)$ is known. In fact the partition function
$Z(\beta)$ for the Hamiltonian (\ref{Ham3}) can be calculated
directly:
\begin{equation}\label{pf2}
Z(\beta)={\int^{+\infty}_{-\infty}}{\prod_{\vec{k}{\neq}0}}d{\theta(\vec{k})}e^{-{\beta}H}=
e^{2{\beta}N}{\prod_{\vec{k}{\neq}0}}{\sqrt{\frac{\pi}{{\beta}f(\vec{k})}}}.
\end{equation}
Then the DOS $g'(E)$ is obtained from the inverse Laplace transformation of  $Z(\beta)e^{-2{\beta}N}$:
\begin{eqnarray}
\nonumber
g{\prime}(E)&=&\frac{1}{2{\pi}i}{\int^{\beta\prime+i\infty}_{\beta\prime-i\infty}}e^{{\beta}E}Z(\beta)e^{-2{\beta}N}d{\beta}\\
\nonumber
            &=&\frac{1}{2{\pi}i}{\int^{\beta\prime+i\infty}_{\beta\prime-i\infty}}e^{{\beta}E}d(\beta){\prod_{\vec{k}{\neq}0}}{\sqrt\frac{\pi}{{\beta}f(\vec{k})}}  \\
\nonumber
            &=&\frac{{\pi}^{\frac{N-1}{2}}}{2{\pi}i}({\prod_{\vec{k}{\neq}0}}{\frac{1}{\sqrt{f(\vec{k})}}}){\int^{\beta\prime+i\infty}_{\beta\prime-i\infty}}{\frac{e^{{\beta}E}}{{\beta}^{\frac{N-1}{2}}}}d{\beta}  \\
\nonumber
            &=&C E^{\frac{N-3}{2}}
\end{eqnarray}
Finally we have:
\begin{equation}\label{dos}
g(E)=C (E+2N)^{\frac{N-3}{2}}
\end{equation}
\begin{equation}\label{const}
C={\frac{\pi^{\frac{N-1}{2}}}{\Gamma(\frac{N-1}{2})}}
{\prod_{\vec{k}{\neq}0}}\frac{1}{\sqrt{f(\vec{k})}}.
\end{equation}
This expression of DOS is the result from the low temperature
approximation of partition function, which is valid only in the
vicinity of the ground state.

The DOS of this model is symmetrical, so the DOS close to the higher
boundary has similar expression. As was pointed out before,
 Wang-Landau algorithm only calculates the
relative value of the density of states. The constant $C$ is of
special use in normalization, which is quite valuable in calculating
the free energy and the entropy. The Eq.~(\ref{dos}) solves the
problem of strong variations of DOS close to the boundary, and at
the same time,  Eq.~(\ref{const}) solves the
problem of normalization. Both are intractable problems when dealing
with continuous systems, and we solve them by the theoretical
derivation. This is a very general method to obtain DOS of a continuous system
close to the ground state.

\begin{figure}[htbp]
\includegraphics[width=0.8\linewidth]{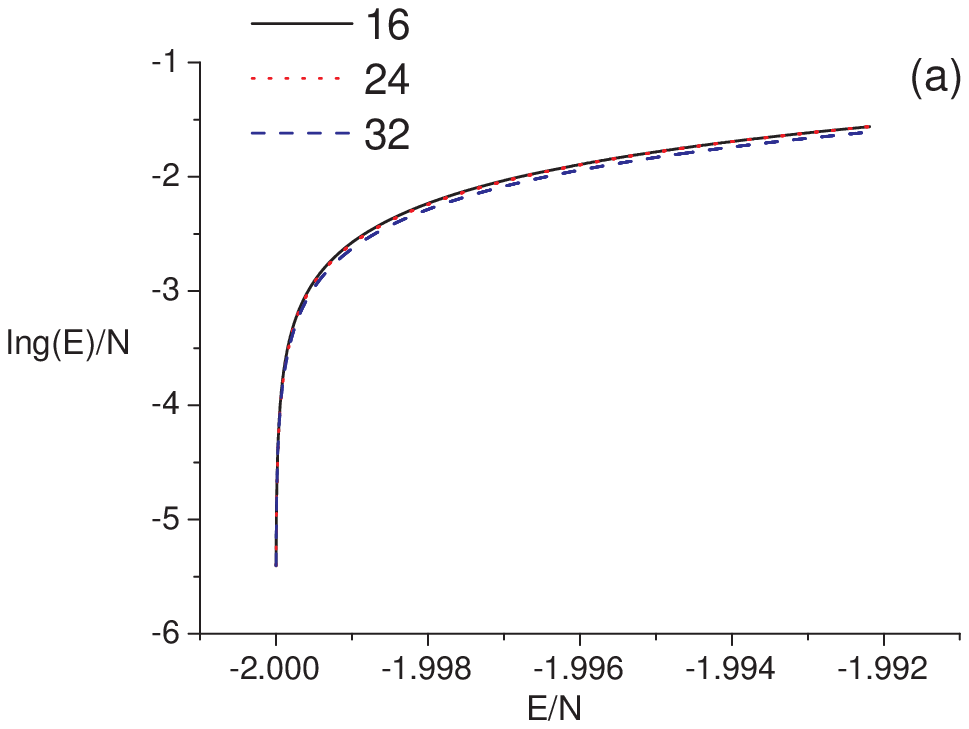} \\
\includegraphics[width=0.8\linewidth]{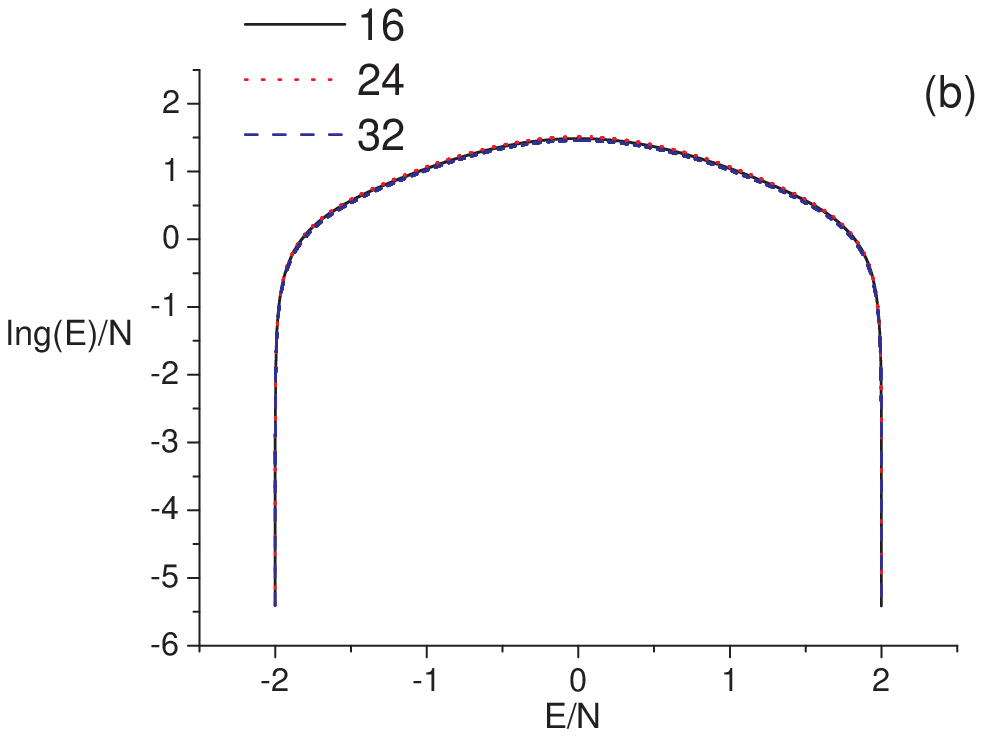}
\caption{(a) the density of states near the lowest energy of two
dimensional XY model on square lattices of size $16\times16$,
$24\times24$, $32\times32$.   (b) the density of states in the whole
energy range of two dimensional XY model on square lattices of size
$16\times16$, $24\times24$, $32\times32$.} \label{fig1}
\end{figure}

\begin{figure}[htbp]
\includegraphics[width=0.8\linewidth]{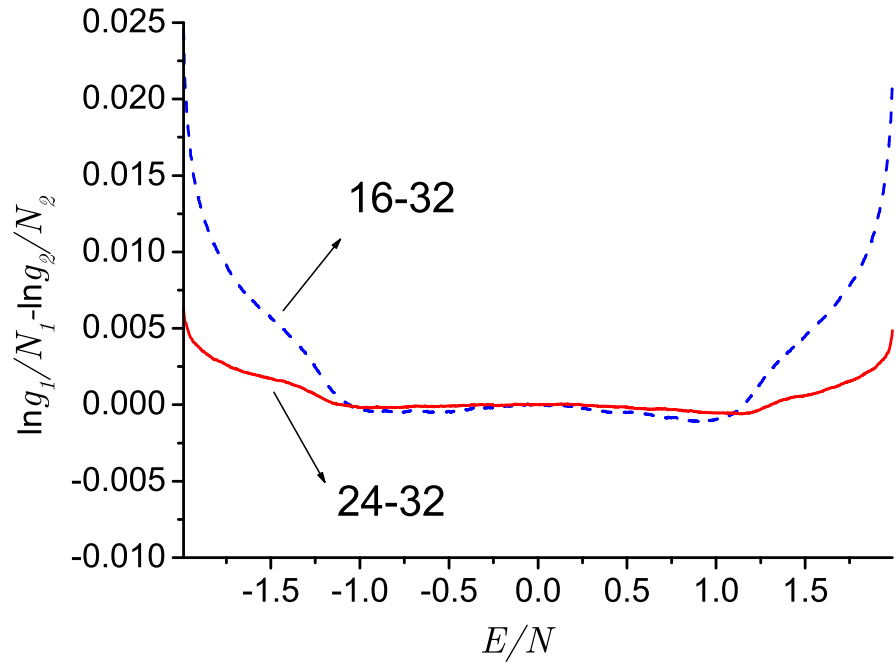}
\caption{The difference of two dimensional XY model on square lattices of size
$16\times16$ and $32\times32$, and $24\times24$ and $32\times32$ after scaling.
The DOS of energy $E=0$ is adjusted so that they have the same value.}
\label{fig2}
\end{figure}

With the above mentioned method,  we simulated the relative DOS for
each enlarged sub-intervals of the middle part, and evaluated the
DOS close to the spectrum boundaries by the analytic expressions.
Since each sub-interval is enlarged a little bit so that there is a
overlap region at each join part, this overlap can help to get a
smooth connect with high accuracy. We also take consideration of the
trick of Shulz et al\cite{BKMD2003PRE} in the accurate calculation
of the DOS on the interval boundaries. The final result of DOS is
shown in Fig. \ref{fig1}. The calculation are performed with finite
systems of $16\times 16$, $24\times 24$ and $32\times 32$, and
periodical boundary condition were used in the simulation. The DOS
shows the generic behavior: $g(E,N){\sim}e^{N{\phi}(E/N)}$ where
${\phi}(E/N)$ is an intensive function. This behavior is consistent
with the scaling law in
ref.\cite{{RRA2002PA},{R2004JSP},{RR1999PRL}}. It should be noted
that this scaling relation is  strictly satisfied only in the
thermodynamic limit where $N \to \infty$. In Fig. \ref{fig2} we show
the small deviations of $\ln g_{16\times16}/N$ and  $\ln
g_{24\times24}/N$ to $\ln g_{32\times32}/N$
 with respect to $E/N$. From the figure we see that the difference between the case of
 $24 \times 24$ and $32 \times 32$ is much smaller than the difference between the case of
 $16 \times 16$ and $32 \times 32$. Based on this result we expect that the case
 of $32\times32$ is a very good approximation
 to the infinite system, and the DOS calculated with $32 \times 32$ should be
 good enough in most applications. However, in the case a phase transition
is occurred the finite size effect may be important and the $32 \times 32$ lattice
may be too small to obtain accurate results as in the case of KT transition.

\subsection{thermodynamic quantities}

With the accurate result of the DOS obtained, some thermodynamic quantities can
be  calculated from it by a simple integration. The internal energy, the specific
heat, the free energy and entropy can be immediately calculated according to the
equations as follows:
\begin{align}
U(T)& =\frac{{\sum_{E}}Eg(E)e^{-{\beta}E}}{{\sum_{E}}g(E)e^{-{\beta}E}}=\langle E
\rangle_T \\ \label{U} C(T)& =\frac{{\partial}U}{{\partial}T}=\frac{\langle
E^2\rangle_T-{\langle E \rangle^2}_T}{T^2}\\ \label{C} Z   &
={\sum_{E}}g(E)^{-{\beta}E}  \\ \label{Z} F(T)& =-T\ln Z   \\  \label{F} S(T)&
=\frac{U(T)-F(T)}{T}  \\  \label{S}\notag
\end{align}
From the expression we see that the partition function can not be
obtained without the knowledge of the
normalization constant $C$, and neither of the free energy and the entropy. When the
temperature $T\rightarrow0$ we may use Eq.~(\ref{dos})  to get the
analytic expressions of thermodynamic quantities at low temperature:
\begin{align}
Z  & ={\int^{-2N}_{2N}}C(E+2N)^{\frac{N-3}{2}}e^{-{\frac{E}{T}}}dE
=C{\Gamma}\left(\frac{N}{2}\right)e^{\frac{2N}{T}}T^{\frac{N-1}{2}}  \\
\label{Zlt} F  & =-T\ln
Z=-T\ln\left[C{\Gamma}\left(\frac{N}{2}\right)\right]-2N-{\frac{N-1}{2}}T\ln T \\
\label{Flt} S  &
=-\frac{{\partial}F}{{\partial}T}=\ln\left[C{\Gamma}\left(\frac{N}{2}\right)\right]
+{\frac{N-1}{2}}\left(\ln T+1 \right)  \\ \label{Slt} C  &
=T\frac{{\partial}S}{{\partial}T}=\frac{N-1}{2}  \\
\label{Clt} U  & =F+TS=-2N+\frac{N-1}{2}T  \\  \label{Ult}\notag
\end{align}

  The low temperature result was obtained in the  assumption that
the total number of lattice sites is very large, that is, the thermodynamic
limit $N\rightarrow\infty$ is assumed. By taking the the zero temperature
limit we get the ground state behavior:
$F\rightarrow -2N$, $S\rightarrow -\infty$, $C\rightarrow \frac{N}{2}$ and $U\rightarrow -2N$.
The behavior of the entropy does not follow the third law
of thermodynamics, which requires that the entropy is zero in the limit.
The reason for this discrepancy is the ignorance of the quantum effect of
our calculation, in fact, in real physical systems the zero temperature
behavior of the system is always quantum. Mathematically the incorrect
zero temperature limit of entropy comes from the finite and continuous density of
states close to the ground state. The zero temperature entropy is given by
 $S(T=0)=\ln g_0$, where $g_0$ is the degeneracy of the ground state.
For XY model, the density of states is continuous so that the number
of states at any specific  point of energy is zero, thus  $g_0=0$ and
$S \to -\infty$.

\begin{figure}[tbp]
\includegraphics[width=0.45\linewidth]{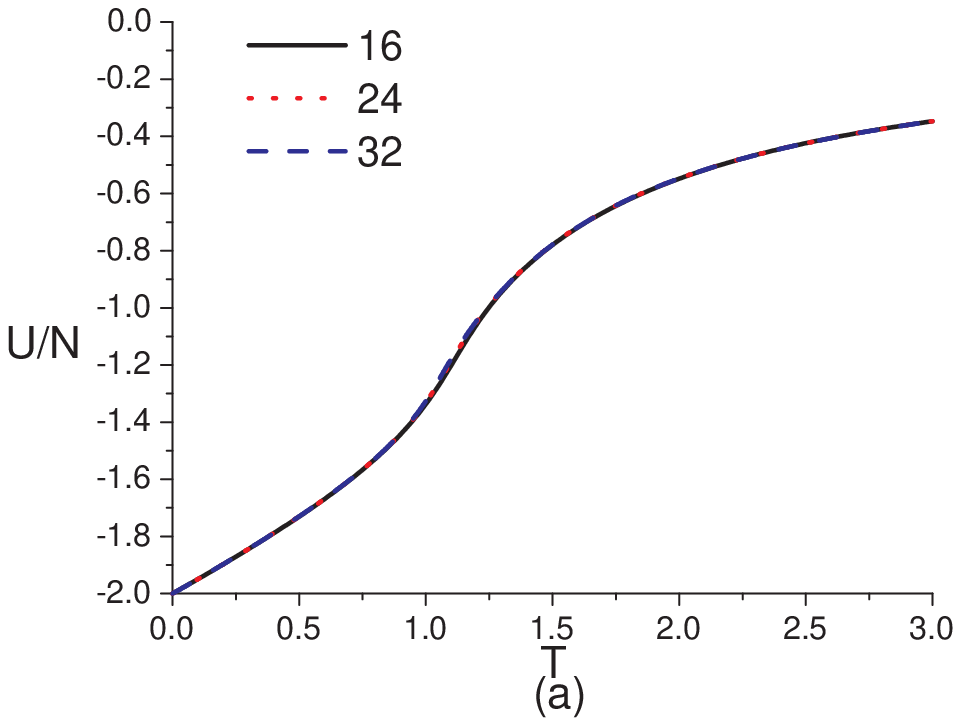} \includegraphics[width=0.45\linewidth]{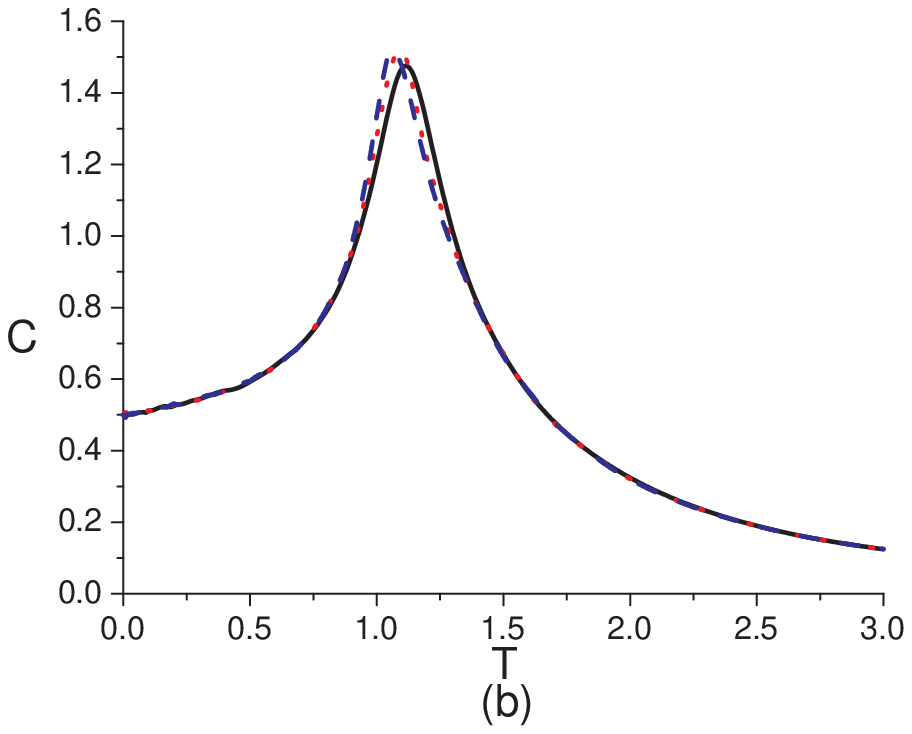}  \\
\includegraphics[width=0.45\linewidth]{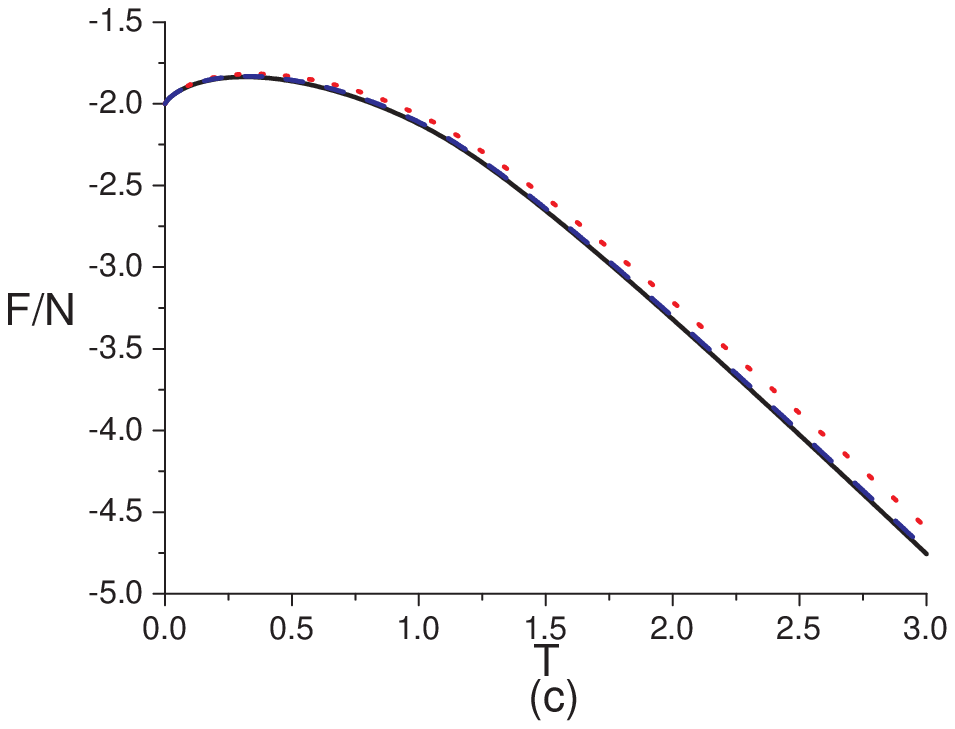} \includegraphics[width=0.45\linewidth]{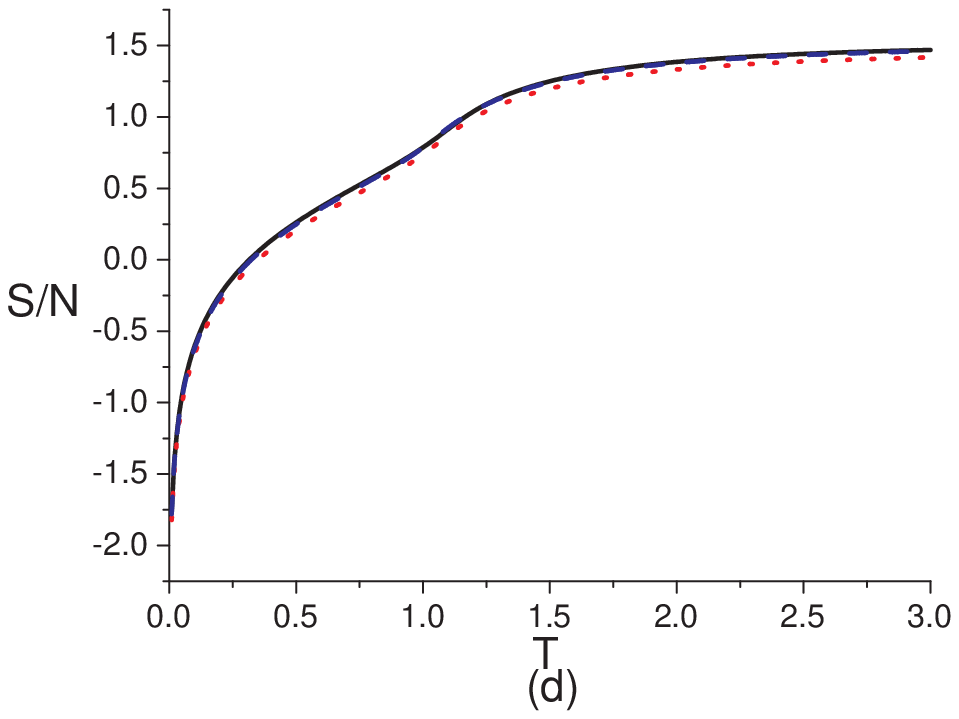}
\caption{ The internal energy, specific heat, free energy and entropy of the two
dimensional  XY model on  square lattices. $16\times16$ solid line, $24\times24$
dot line and $32\times32$ dash line.} \label{fig3}
\end{figure}

The results of the thermodynamical quantities in a large temperature range was
evaluated by numerical integration and are shown in Fig. \ref{fig3}. The figures
plot the quantities per site for different sizes of the system, and it is clear
from the figures that the results of different sizes of the lattices basically
follow the same curve, which indicates that the size effect is not quite
important for system size larger than $16 \times 16$, the smallest system we
calculated.

As we only consider short range interaction, this result is
consistent with ref.\cite{{RRA2002PA},{R2004JSP},{RR1999PRL}}. The
peak of the specific heat differs with the size of the system. In
the temperature near $T=1$, the entropy shows a sudden rise, which
indicates the rise of the number of microscopic configurations. We
know in the temperature of KT transition the vortex pairs break up
and the restrict is released\cite{JV1979PRB}, which also gives a
rise to the number of microscopic configurations. So the abnormal
behavior of the entropy curve may probably show the KT transition.
As KT transition is very weak, it can hardly be observed from the
free energy curve. It is known that the effective vortex pair
interaction is long ranged so that our current system size is
too small to describe the KT transition properly.
 Thus the accurate transition temperature is yet not
determined by this method.

\section{CONCLUSIONS}
\label{sect3}

We calculated the DOS of two dimensional XY model on a square lattice
by using the
Wang-Landau algorithm combined with analytical expressions on the spectrum
boundaries. And from the DOS we calculated the internal
energy, the specific heat, the free energy and the entropy. From the
curve of entropy we see some symptoms of phase transition. We find
that simulation algorithms will meet difficulties in the
calculation of DOS close to the spectrum boundaries with
continuous models. We propose a general method to solve such difficulty
by using the analytical expressions at the boundaries.

This work is supported by the National Nature Science Foundation of China under grant
\#10334020 and \#90103035 and in part by the National Minister of Education
Program for Changjiang Scholars and Innovative Research Team in University.

\newpage

\end{document}